\documentclass[5p]{elsarticle}

\usepackage{lineno}
\usepackage[pdfauthor=Placeholder]{hyperref}
\usepackage{amsmath}
\usepackage{amsfonts}
\usepackage{graphicx}
\usepackage{subfigure}
\usepackage{xcolor}
\usepackage{comment}
\usepackage{graphics}
\usepackage{subfigure}
\usepackage{url}
\usepackage{diagbox}
\usepackage{ulem}
\usepackage{import}

\modulolinenumbers[1]

\definecolor{rred}{rgb}{0.8,0.1,0.1}
\definecolor{ggreen}{rgb}{0.1,0.7,0.1}
\definecolor{bblue}{rgb}{0.1,0.1,0.7}

\begin{document}

\begin{frontmatter}

    \title{Fast Muon Tracking with Machine Learning Implemented in FPGA}

    \author[mymainaddress]{Chang Sun}

    \author[mymainaddress]{Takumi Nakajima\corref{mycofirstauthor}}
    \cortext[mycofirstauthor]{Co-first author}

    \author[mymainaddress]{Yuki Mitsumori}

    \author[mymainaddress]{Yasuyuki Horii\corref{mycorrespondingauthor}}
    \cortext[mycorrespondingauthor]{Corresponding author}
    \ead{yhorii@hepl.phys.nagoya-u.ac.jp}

    \author[mymainaddress,mysecondaryaddress]{Makoto Tomoto}

    \address[mymainaddress]{Nagoya University, Chikusa-ku, Nagoya 464-8602, Japan}
    \address[mysecondaryaddress]{High Energy Accelerator Research Organisation (KEK), Oho, Tsukuba 305-0801, Japan}

    \begin{abstract}
        In this work, we present a new approach for fast tracking
        on multiwire proportional chambers with neural networks.
        The tracking networks are developed and adapted
        for the first-level trigger at hadron collider experiments.
        We use Monte Carlo samples generated by Geant4
        with a custom muon chamber, which resembles part of the thin gap chambers
        from the ATLAS experiment, for training and performance evaluations.
        The chamber has a total of seven gas gaps,
        where the first and last gas gaps are displaced by $\sim 1.5$~m.
        Each gas gap has 50~channels with a size of 18--20~mm.
        Two neural network models are developed and presented:
        a convolutional neural network and a neural network
        optimized for the detector configuration of this study.
        In the latter network, a convolution layer is provided for each of three groups
        formed from 2--3 gas gaps of the chamber,
        and the outputs are fed into multilayer perceptrons in sequence.
        Both networks are transformed into hardware description language
        and implemented in Virtex UltraScale+ FPGA.
        The angular resolution is 2~mrad, which is comparable
        to the maximum resolution of the detector estimated by the minimum $\chi^2$ method.
        The latency achieved by the implemented firmware is less than 100~ns,
        and the throughput rate is 160~MHz.
    \end{abstract}

    \begin{keyword}
        Muon Tracking, Machine Learning, Field-Programmable Gate Array, Artificial Neural Network
    \end{keyword}

\end{frontmatter}


\section{Introduction}

Triggering muons is critically important
in proton-proton collider experiments.
For instance, in 2012, a new particle with a mass of approximately 125~GeV
was observed at the ATLAS and CMS experiments~\cite{ATLAS_Higgs, CMS_Higgs}
using a muon trigger~\cite{ATLAS_muon_trigger, CMS_l1_trigger}
at the CERN Large Hadron Collider (LHC).
Subsequent measurements indicate that the new particle
is consistent with the Higgs boson
in the standard model~\cite{ATLAS_Higgs10, CMS_Higgs10}.

An upgrade to the LHC, which is known as the High-Luminosity LHC (HL-LHC),
is expected to become operational in 2027~\cite{hl-lhc}.
The peak luminosity of the HL-LHC
would ultimately be increased to $7.5\times 10^{34}~{\rm cm}^{-2}{\rm s}^{-1}$,
which is 7.5 times the design luminosity of the LHC ($1\times 10^{34}~{\rm cm}^{-2}{\rm s}^{-1}$).
The trigger and data-acquisition system of the ATLAS experiment requires upgrades~\cite{tdaq-tdr}
to accommodate the increase in luminosity.
In the original first-level muon trigger, only selected and processed hit data
of the muon spectrometer are transferred
from the frontend to the backend electronics.
However, the new system will send all hit data
to the backend electronics,
exploiting advances in data-transfer technology,
and all hits will be processed
using high-end FPGAs integrated into the backend electronics.

The ability of backend FPGAs to retrieve full hit data
provides a unique opportunity to boost the first-level muon trigger's performance.
A throughput of $\geq 40~{\rm MHz}$ and
a latency within ${\cal O}(100~{\rm ns})$ are required.
One method proposed for the new system
is a fast tracking algorithm known as ``pattern matching''~\cite{tdaq-tdr},
which compares the received hit data
with predefined hit lists with corresponding track information assigned.
Another possibility,
which is the subject of this paper,
is to use a neural network to perform fast tracking,
exploiting the augmented on-board computing power.
Here, the hit data are used as the neural network's input,
and the track information is outputted.

Machine learning techniques are rapidly being developed for
and applied in high-energy physics experiments.
A living review is provided in Ref.~\cite{ML_summary}.
Refs.~\cite{Duarte_2018, Summers_2020, Ngadiuba_2020, qkeras}
show how different architectures
can be deployed on a first-level trigger using the hls4ml library.
These works have opened a new era
for using machine learning for trigger systems.
Novel fast tracking on a silicon detector
has been and continues to be developed~\cite{tdaq-tdr, CMS-PhaseII-L1-TDR, FTK, CDF_SVT};
however, the use of machine learning is challenging
because of the large number of channels.
A muon chamber has fewer channels,
and fast muon tracking with machine learning
has been attempted~\cite{ML_muon_CMS}.

In this study, a novel approach of using neural networks
for online muon tracking is introduced.
Neural network models are developed
for the first-level muon trigger at high-luminosity hadron colliders.
Two neural network models are implemented in Xilinx Virtex UltraScale+ FPGA
and are tested on Monte Carlo (MC) samples
through Vivado post-implementation simulations~\cite{vivado}.
The tracking performance, latency, and required FPGA resources
are evaluated with different hyper parameters.
For the tracking performance, focus is placed on the angular resolution,
which is critical for the momentum determination
(and hence the trigger rate) at the first-level muon trigger~\cite{tdaq-tdr}.
The dependence of the tracking performance
on the detector noise level is also studied.

The remainder of this paper is organized as follows.
The model for the MC sample
as well as the tracking performance by a conventional minimum $\chi^2$ method
is described in Section~\ref{sec:sim}.
The workflow and software are introduced in Section~\ref{sec:workflow}.
The network design and evaluations
for the two types of networks are described
in Sections~\ref{sec:cnn} and \ref{sec:mnn}.
The conclusion is provided in Section~\ref{sec:conclusion}.

\section{Simulation}
\label{sec:sim}

\subsection{Simulation samples}
\label{sec:simulation_samples}

\begin{figure}[htbp]
    \centering
    \includegraphics[width=0.90\columnwidth]{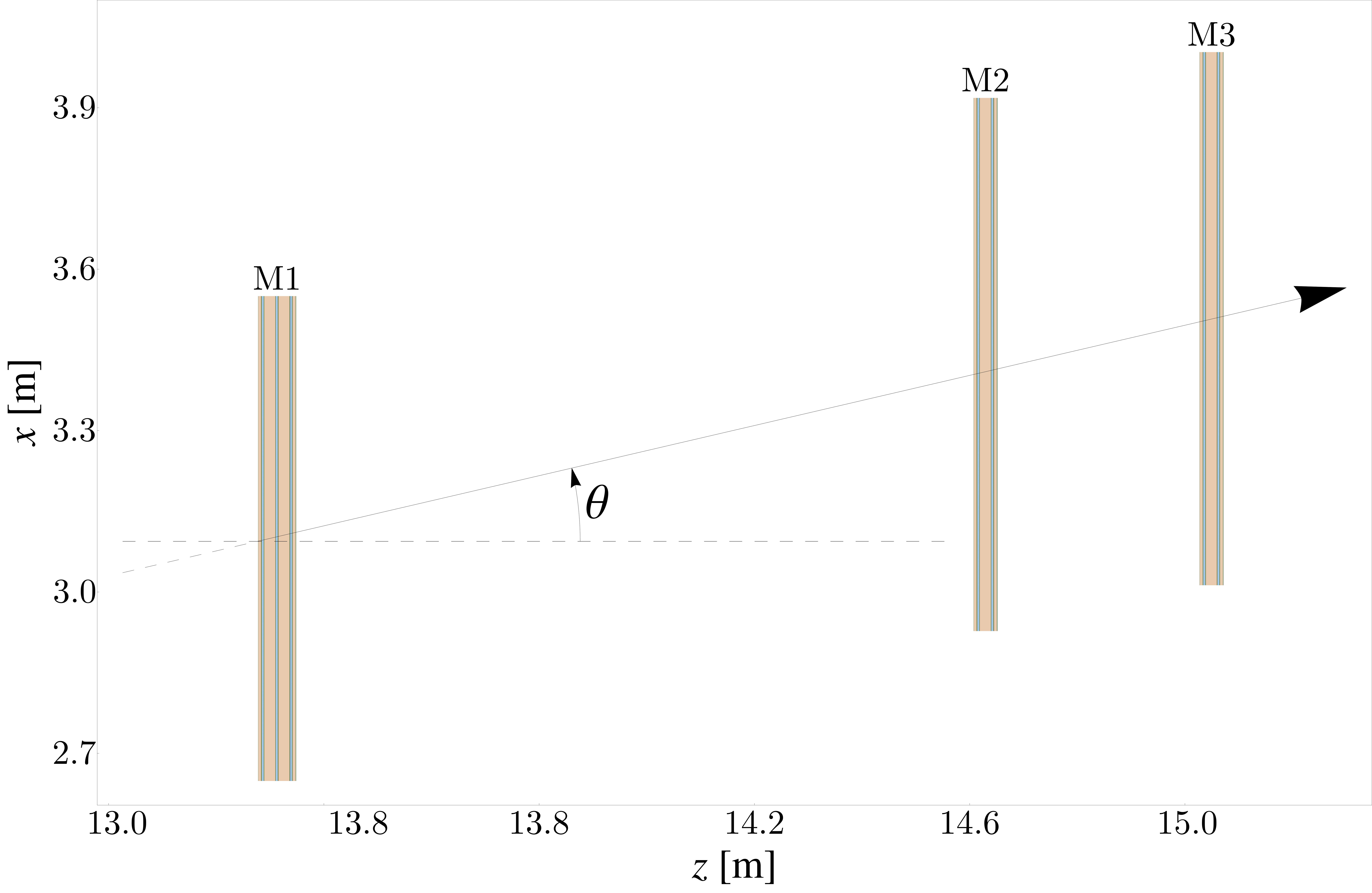}
    \caption{Schematic of the detector model.
        The $z$ coordinates of the chambers
        are identical to those of TGC detectors in the ATLAS experiment.
        The three plates---M1, M2, and M3
        in ascending $z$-position order---are aligned
        with their normals parallel to the $z$ axis.
        The arrow represents an incoming muon,
        and $\theta$ is defined as the polar angle.}
    \label{detector_schematic}
\end{figure}

\begin{figure}[htbp]
    \centering
    \includegraphics[width=0.90\columnwidth]{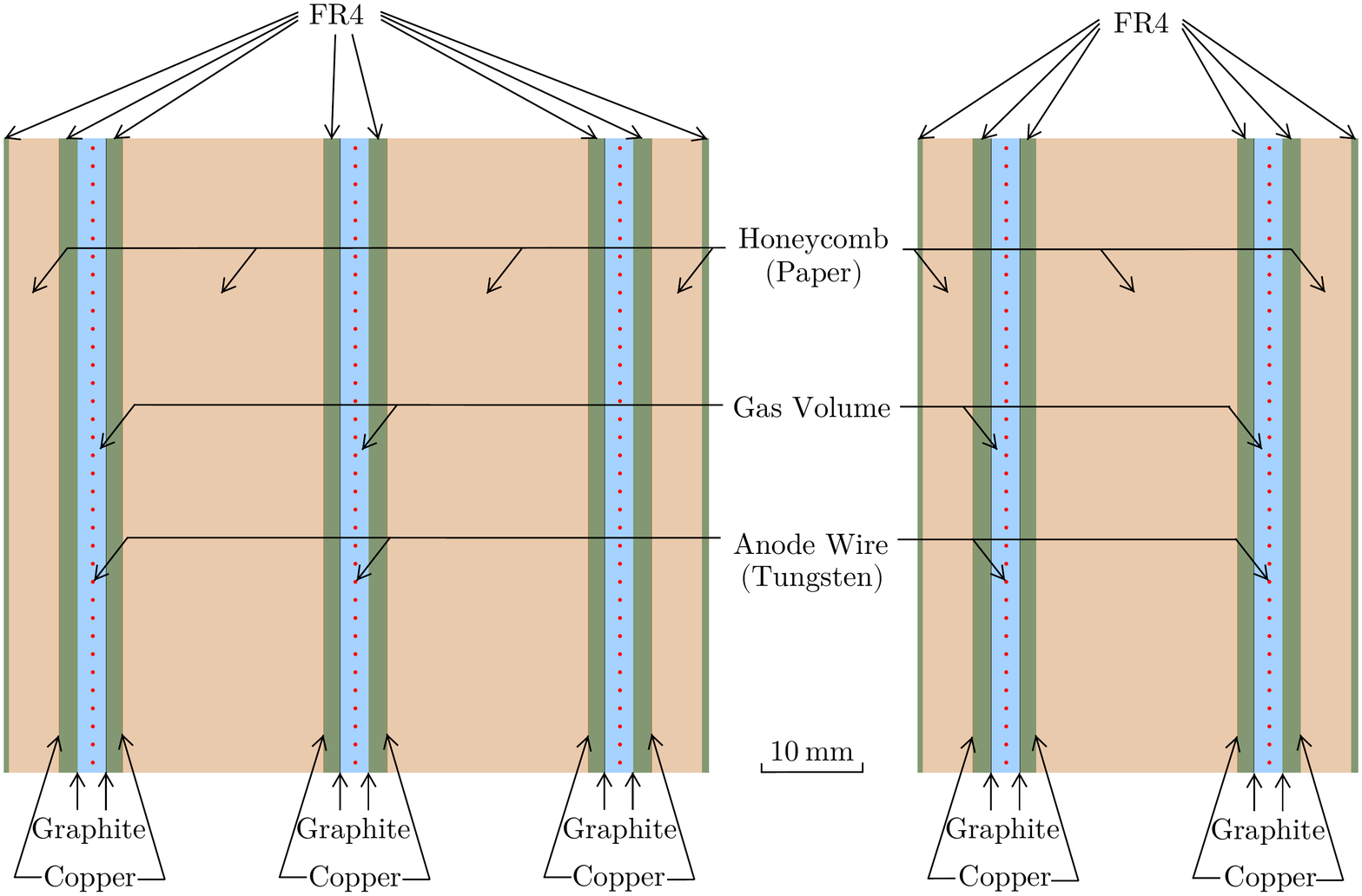}
    \caption{Cross-sectional view of the triplet (left) and doublet (right) plates.}
    \label{detector_cross_section}
\end{figure}

The networks presented in this work were
trained and evaluated on events
simulated with Geant4~\cite{G4-short}.
In the simulation, a gas chamber detector shape
similar to the thin gap chamber (TGC)
of the ATLAS experiment~\cite{tgc_structure}
was formed with an approximate pseudorapidity ($\eta$) coverage of $2.03\le|\eta|\le 2.26$.
A full schematic of the detector model used
is shown in Fig.~\ref{detector_schematic}.
Rectangular plates were used instead of the trapezoidal plates
of the TGC detector for simplicity.
The angular resolution of the tracking
shown in this paper is independent
from the simplifications of the detector geometry
because the wire orientations are identical.
Three plates---one triplet and two doublet plates---were used in the simulation.
The triplet plate corresponds to the TGC's innermost plate
and has dimensions of $900\text{ mm}\times 500\text{ mm}\times 70\text{ mm}$.
The two doublet plates correspond to the TGC's outer plates
and have dimensions of
$990 \text{ mm} \times 500 \text{ mm} \times 44 \text{ mm}$.
The triplet plate is named M1,
and the doublet plates are named M2 and M3.
All the plates were positioned with their central points
set to the TGC plates' central positions.
The $z$ positions\footnote{
    The $z$ axis is taken as the proton beam axis in the ATLAS experiment,
    and is taken accordingly in this study.}
of the M1, M2, and M3 plates were identical to TGC's plates
(Fig.~\ref{detector_cross_section}).
The wire pitch for all the plates was $1.8$~mm,
and 10 (11) wires were grouped into channels
for triplet (doublet) plates.
No magnetic field was introduced
because the TGC detector is located outside the toroidal magnet
at the ATLAS experiment.

The muon source was placed in front of the M1 plate at $z=13.3$~m.
Multiple scattering and emissions
before arriving at the muon chamber were ignored because
the intrinsic performance of the tracking with the muon chamber
was the topic of interest.
At the ATLAS experiment, muons with transverse momentum $p_{\rm T} \ge 20$~GeV
are deflected, at most, $\pm 30$~mrad
from the trajectory of muons
with infinite momentum when impacting the TGC detector.
The sign of the deflection angle depends
on the electric charge of the muon.
In this study, muons' direction was uniformly distributed
in the range $\pm 30$~mrad from the angle of the straight track,
which was independently uniformly distributed on the polar angle $\theta$.
The straight tracks were aligned directly from the interaction point.
A single muon was generated for each event.
Tracking of the minimum ionizing particle is critical
for muon triggerring at the ATLAS experiment, and
all muons were generated with $p_{\rm T} = 20$~GeV in this study.
All data associated with tracks passing through all seven gas gaps
were collected for both training and performance evaluation.
A channel generated a readout value of 1 if it was fired and 0 if not.
The fired channels are referred to as ``hits'' in this work.

With the aforementioned settings, $3 \times 10^6$ events
were generated for this study.
Twenty percent of the data was used as the test set,
which was not exposed to any models introduced in this study
prior to the performance analysis.

To achieve a more realistic background environment,
we prepared extra simulation samples with artificial noises.
These samples are generated by randomly setting the readout value to $1$
with probability $p$ for all channels independently of the generated dataset.
Hereafter, $p$ is represented as the ``noise level''.

\begin{figure*}[htbp]
    \centering
    \includegraphics[width=0.95\textwidth]{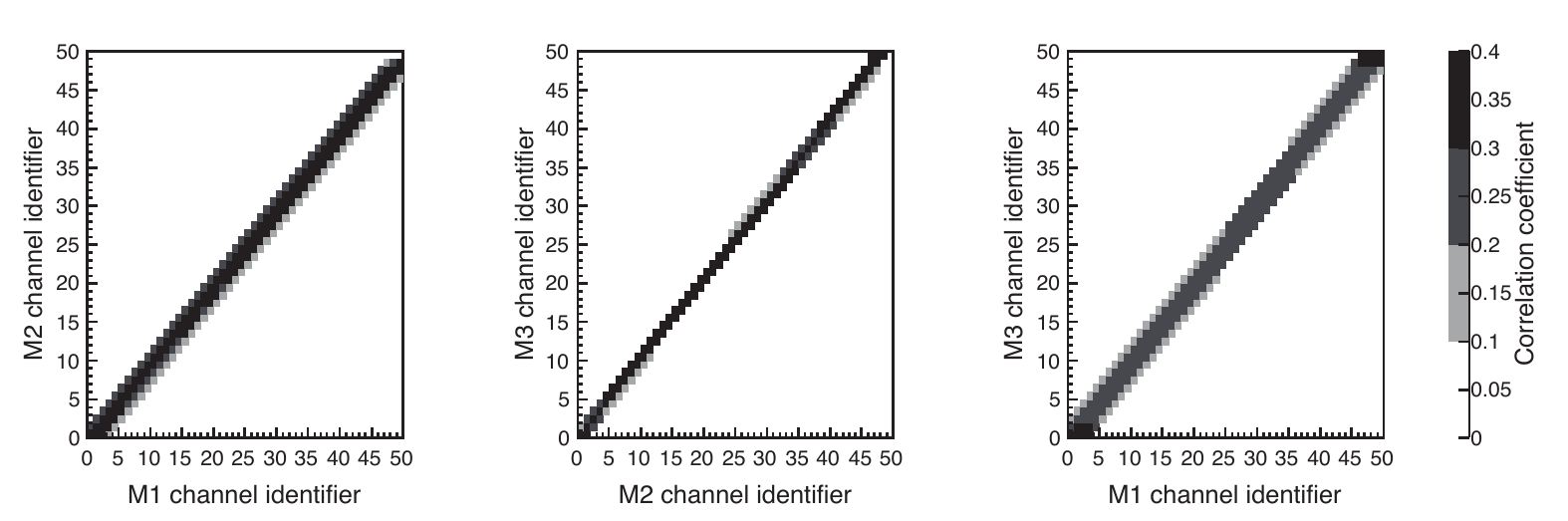}
    \caption{Channel-wise correlation matrices
    for combinations of the M1, M2, and M3 plates.}
    \label{data_corr_mat}
\end{figure*}

The channel-wise correlation matrices of the channel outputs
between the plates are shown in Fig.~\ref{data_corr_mat}.
The elements of the correlation matrices,
the correlation coefficients $\rho_{i,j}$, are defined as
$\rho_{i,j} \equiv E[\{x_i - E(x_i)\}\{x_j - E(x_j)\}] /$
$\sqrt{ E[ \{x_i - E(x_i)\}^2 ] E[ \{x_j - E(x_j)\}^2 ] }$,
where $x$ is $1$ if the corresponding channel has a hit and $0$ otherwise,
$i$ and $j$ are the channel identifiers for the given two plates,
and $E$ represents the expectation value.
These correlation matrices were used for determining some of the hyper parameters
and for reducing the number of multiplication operations required by the network.

\subsection{Reference tracking performance}

For reference, the best possible resolution of this detector was estimated
using the minimum $\chi^2$ method.
In each event, one straight line was randomly placed in the detector's volume
according to the distribution of muon trajectories,
as described in Section~\ref{sec:simulation_samples},
and all channels passed by the line were marked as fired.
Because there was no false hit, an ordinary $\chi^2$ fit was used
to estimate the resolution.
Specifically, when a gas gap had exactly one hit,
the coordinate of the center of the channel was used for the fit
with an error $\sigma$; if a gas gap had two adjacent hits,
the coordinate of the center of the two channels
combined with an error $\sigma/\sqrt 2$ was used for the fit.
This method is referred to as ``ideal $\chi^2$'' in this work.
The detector's angular resolution determined by this method is 1.7~mrad.

The performance determined from a $\chi^2$-fitting method iterating
through all viable hit combinations
was evaluated for comparison to the network in this study
on the simulated events.
The exact algorithm used for this fit is described as follows:
1) One hit was taken from each gas gap with non-zero hits,
and a simple $\chi^2$ fit was performed for all combinations.
2) The fitted track with the lowest $\chi^2$ score was selected as the output.
Here, given the uniform channel sizes,
the hits from the same plate were regarded as having the same uncertainty.
This method is referred to as ``iterative $\chi^2$'' in this work.
The angular resolution obtained in this manner for the simulated events is 2.2~mrad.
Although the iterative $\chi^2$ method is more resistant to noises,
it uses a maximum of one hit per gas gap
and does not fully exploit the position information for a muon
that provides hits in two neighboring channels.

\section{Workflow}
\label{sec:workflow}

All neural networks used in this study were first
defined and trained using Keras 2.4.3~\cite{keras} and
QKeras 0.8.0~\cite{qkeras}
with TensorFlow 2.2.0~\cite{tensorflow2015-whitepaper} as the backend.
The networks were then translated into an intermediate C++ script
using hls4ml 0.5.0~\cite{Duarte_2018}
with the setup of \texttt{ReuseFactor = 1},
\texttt{Strategy = Latency}, and \texttt{IOType = io\_parallel}.
The synthesis was performed with Vivado HLS 2020.1~\cite{vivado_hls}.
The implementation was carried out with Vivado 2020.1~\cite{vivado},
targeting a Xilinx Virtex UltraScale+ XCVU13P FPGA\footnote{XCVU13P-L2FHGA2104E}
with a clock frequency of 160~MHz.
The performance was evaluated with a post-implementation
functional simulation running on the Vivado software.

\section{Muon tracking with convolutional neural networks}
\label{sec:cnn}

\subsection{Network design}

A convolutional neural network (CNN) is a class of deep neural network,
widely applied to the analyses of visual images.
In this study, the detector outputs were considered
to be visual images and muon tracks were reconstructed by a CNN.
The input of the network was the hit information for all channels $\langle 50 \times 7 \rangle$
and the output was the muon track angle~$\theta$.

\begin{figure*}[htbp]
    \centering
    \subfigure[\label{fig:CNN_big} Large-scale CNN model (\textbf{SW})]
    {\includegraphics[width=0.95\textwidth,pagebox=cropbox,clip]{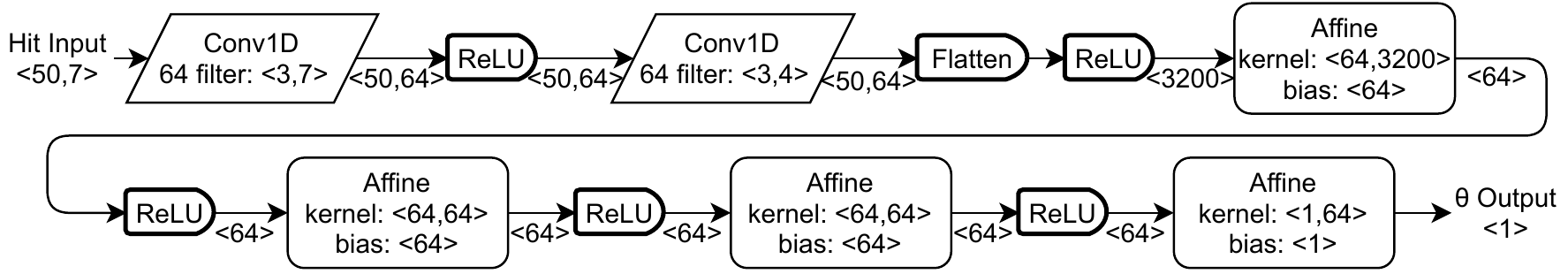}}
    \subfigure[\label{fig:CNN_small} Compact CNN model (\textbf{BL}, \textbf{QF8}, \textbf{QF6})]
    {\includegraphics[width=0.95\textwidth,pagebox=cropbox,clip]{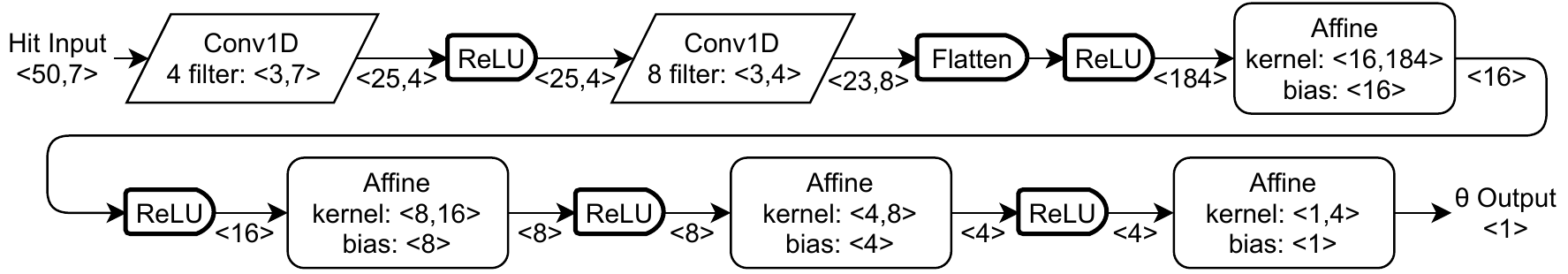}}
    \caption{Schematics of the large-scale and compact CNN models.
    Both are composed of two convolution (``\texttt{Conv1D}'') layers
    and four affine (``\texttt{Affine}'') layers.
    The kernel size for the \texttt{Affine} layers refers to
    $\langle$\texttt{output dimension}, \texttt{input dimension}$\rangle$.
    The kernel size for \texttt{Conv1D} layers refers to
    \texttt{filter size} and $\langle$\texttt{input channel}, \texttt{filter count}$\rangle$.
    The \texttt{<*>} notation on arrows
    indicates that the array shape passed through the arrow.}
\end{figure*}

A large-scale network that maximizes the muon tracking performance was designed;
its schematic is shown in Fig.~\ref{fig:CNN_big}.
This design was primarily inspired by the VGG network~\cite{VGG}.
Multiple convolution layers extract the features of the muon track,
and the affine (also known as fully connected) layers
located after the convolution layers
extract the muon track angle from the extracted features.
This network is denoted as software (\textbf{SW}).

A compact version of the network
was designed by downscaling the \textbf{SW} network
with the objective of reducing FPGA resource utilization
and achieving a smaller latency value.
This network is referred to as baseline (\textbf{BL}),
and its schematic is shown in Fig.~\ref{fig:CNN_small}.

The filter sizes of convolution layers were chosen
to cover almost all hits from a single muon track.
The exact numbers were determined
by the channel-wise correlation matrix shown in Fig.~\ref{data_corr_mat}.

Quantization was performed on the compact network
to further reduce FPGA resource utilization and latency.
All parameters, including weights, biases, and intermediate values, in the quantized networks
were quantized to \texttt{ap\_fixed<$\cdot,\cdot$>} or \texttt{ap\_ufixed<$\cdot,\cdot$>}.
Overflow and round behaviors were set to be default with wraparound (\texttt{AP\_WRAP})
and truncation to minus infinity (\texttt{AP\_TRN}).
The numbers of integer bits assigned to each layer's weights,
biases, and intermediate values were selected to be the minimum number
such that no overflow would occur for the entire training set.
By contrast, the number of float bits assigned ($n$) was fixed for all parameters
except accumulators, which had $(n+2)$ float bits instead of $n$.
As the only exception to this rule, the output layer always had $9$ unsigned integer bits and $7$ float bits.
A network quantized with $n$ float bits assigned is denoted as \textbf{QF[n]}.
Here, $n = 8$ and $6$ were used for the compact network,
and the resultant quantized compact networks are
referred to as \textbf{QF8} and \textbf{QF6}, respectively.
Because no significant performance difference was observed
between post-training quantization and quantization-aware training,
only post-training quantization was performed.

All the networks were fully pipelined and
able to accept an event
every $1/(160~{\rm MHz}) = 6.25~{\rm ns}$.
No initialization was required
between the processing of consecutive events,
which is a key feature of the designed network
to be used at the first-level trigger at hadron-collider experiments.

\subsection{Performance}

\begin{figure}[htbp]
    \centering
    \includegraphics[width=0.90\columnwidth]{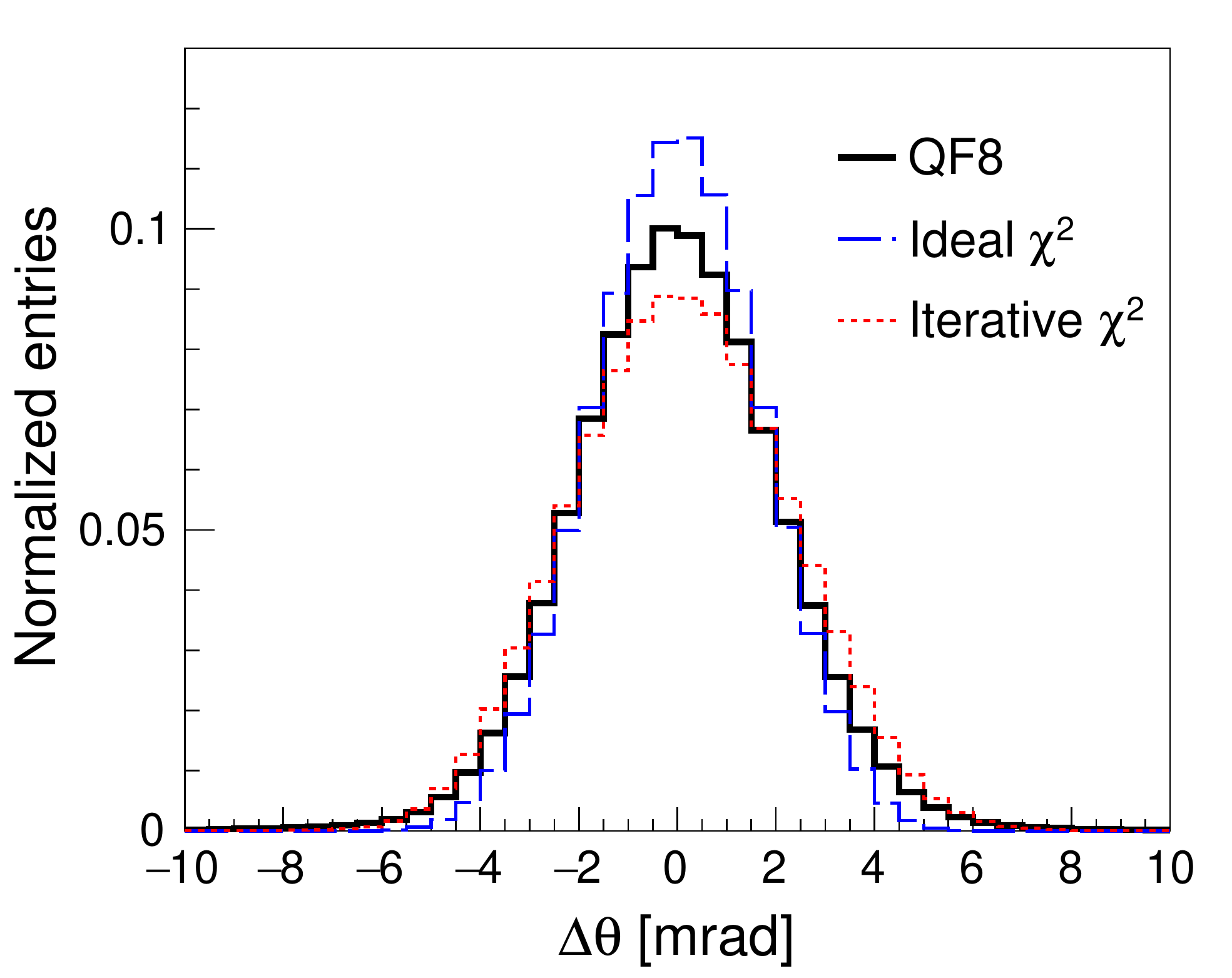}
    \caption{Distributions of the angle difference between the MC truth track segment
        and the reconstructed track segment.
        The solid (black) histogram shows the distribution for the track segments
        reconstructed by the \textbf{QF8} network.
        The dashed (blue) and dotted (red) histograms show the distributions
        for track segments reconstructed by the ideal and iterative $\chi^2$ methods, respectively.}
    \label{fig:software_theta}
\end{figure}

The angular resolution was evaluated from the widths of the distributions
of the difference between the CNN output and the MC truth angle.
Figure~\ref{fig:software_theta} shows the distributions for the \textbf{QF8} network.
The distributions of the ideal and iterative $\chi^2$ methods are also shown for reference.
Table~\ref{tab:fpga-table} shows the values of the angular resolution
for the networks examined in this study.
The \textbf{SW} network provides the best angular resolution of 1.7~mrad.
The angular resolution of the \textbf{BL},
\textbf{QF8}, and \textbf{QF6} networks
ranges from 2.0 to 3.4~mrad.

\begin{table*}[hbtp]
    \caption{The angular resolution, latency,
    and resource utilization for CNN examined in this study.
        The numbers in parentheses show
        the rates of the resource utilization per super logic region.
        The angular resolutions for \textbf{SW} and \textbf{BL}
        are estimated using software,
        whereas those for \textbf{QF8} and \textbf{QF6}
        are estimated using a Vivado post-implementation simulation.
        The \textbf{SW} and \textbf{BL} networks
        are provided without the intention to be run on the hardware;
        thus, the values of the latency and resource utilization are omitted.}
    \label{tab:fpga-table}
    \centering
    \resizebox{\textwidth}{!}{
    \begin{tabular}{lcccccc}
        \hline
        Model         & Resolution~[mrad] & Latency~[ns] & DSP48         & LUT            & FF           & BRAM    \\
        \hline \hline
        \textbf{SW}   & 1.7               & -            & -             & -              & -            & -       \\
        \textbf{BL}   & 2.0               & -            & -             & -              & -            & -       \\
        \textbf{QF8}  & 2.2               & 81           & 2087~(68\%)  & 66,441~(15\%)  & 19,849~(2\%) & 0~(0\%) \\
        \textbf{QF6}  & 3.4               & 81           & 607~(20\%)    & 61,749~(14\%)  & 11,702~(1\%) & 0~(0\%) \\
        \hline
    \end{tabular}
    }
\end{table*}

The latency and the resource utilization were obtained for the quantized networks
and are summarized in Table~\ref{tab:fpga-table}.
Both the \textbf{QF8} and \textbf{QF6} networks were implemented on Vivado
without timing violation
and achieved a latency less than 100~ns.

\section{Muon tracking with multistage neural networks}
\label{sec:mnn}

\subsection{Network design}

\begin{figure*}[htbp]
    \centering
    \includegraphics[width=0.95\textwidth]{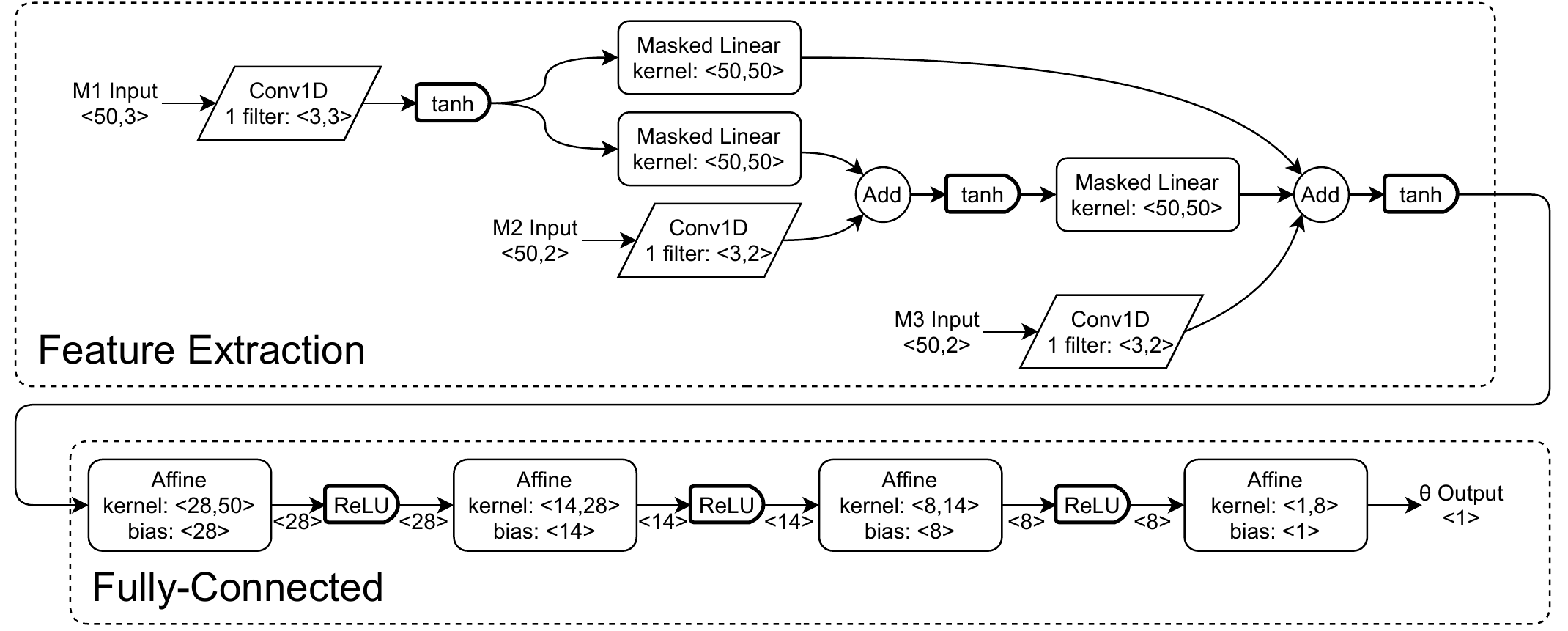}
    \caption{Schematic of the neural network
    optimized for the detector configuration.
    The kernel size for the \texttt{Affine} layers refers to
    $\langle$\texttt{output dimension}, \texttt{input dimension}$\rangle$.
    The kernel size for the \texttt{Conv1D} layers refers to
    \texttt{filter size} and $\langle$\texttt{input channel}, \texttt{filter count}$\rangle$.
    The \texttt{<*>} notation on arrows
    indicates that the array shape passed through the arrow.
    Arrows without an array shape specified are passing arrays of shape 50.}
    \label{rnn_sche}
\end{figure*}

We designed another network to take maximum advantage
of the strong correlation of the hits between plates of the detector.
A schematic of the network is shown in Fig.~\ref{rnn_sche}.
The hit information on the channels from plates M1, M2, and M3
were fed into the network separately in the form of 1-bit arrays of shapes
$\langle 50\times 3 \rangle$, $\langle 50\times 2 \rangle$,
and $\langle 50\times 2 \rangle$, respectively.
The network consists of two building blocks: a feature-extraction network
that exploits the layered structure of the detectors
and a generic fully-connected net that outputs the final values.

The design of the feature extraction part was inspired
by the track-following scheme~\cite{Hennequin_2020},
where multiple hits are combined together
by detector stations and added sequentially
in the inside-out direction to form the full track.
The convolution layers extract real vectors encoding hit-position information from each plate,
whereas the ``masked linear'' layers (i.e., highly sparse affine layers)
define search windows and project the track information from one plate to the next as real vectors.
The previous track information (i.e., the output of the last masked linear layer)
is merged with the current observation (i.e., the output of the convolution layer)
via a simple addition followed by a hyperbolic tangent activation function
to allow some inefficiency for the detector while still favoring the track with more hits.

\begin{figure}[htbp]
    \centering
    \includegraphics[width=0.90\columnwidth]{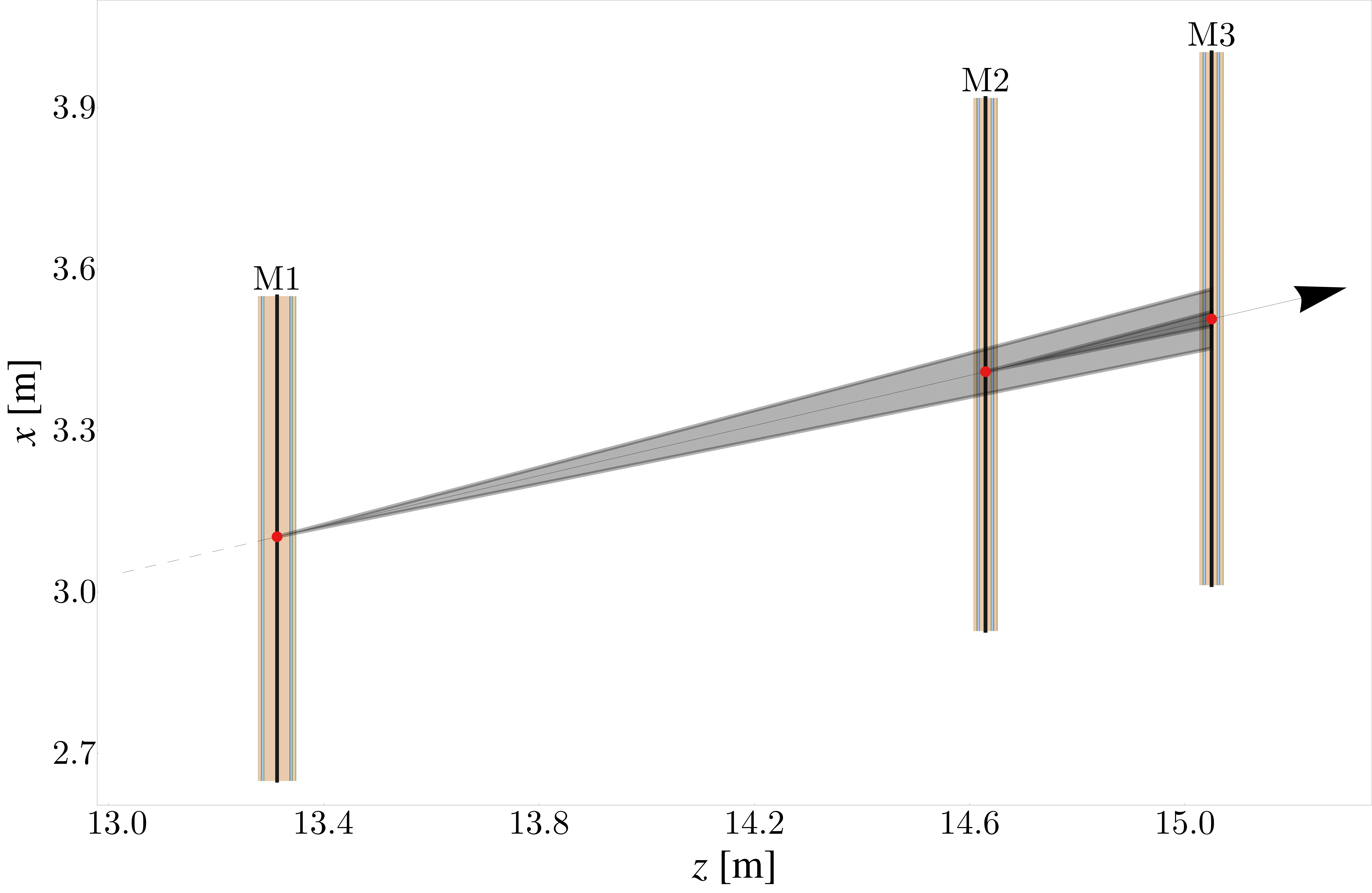}
    \caption{Demonstration of how the mask matrices work. The three black lines on the plates represent the observation and state vectors' position. The gray shaded cones represent the search region for a single hit on M1 or M2 marked red.}
    \label{nn_demo}
\end{figure}

Channel-wise correlation matrices between the plates
(Fig.~\ref{data_corr_mat})
were used to reduce the number of multiplication operations
required by the networks.
We defined the search windows encoded in the masked linear layers
by requiring a correlation coefficient greater than $0.01$.
All weights outside the windows were constrained to be zero.
This practice was responsible for the highly sparse nature
of the masked linear layers,
where the sparsity was $\sim 90\%$.
A graphical explanation
of how the masked linear layers work is provided in Fig.~\ref{nn_demo}.

The network with full precision in the software simulation is
referred to as the \textbf{BL} network.
Three quantized networks were built from the \textbf{BL} network
following the procedure specified in Section~\ref{sec:cnn},
except that, after quantization was applied to the trained \textbf{BL} network,
further quantization-aware training was performed
to mitigate the performance drop.
These networks were quantized with 7, 5, and 3 float bits and are referred to
as \textbf{QF7}, \textbf{QF5}, and \textbf{QF3}, respectively.
All the networks were fully pipelined,
like the networks described in Section~\ref{sec:cnn},
and were able to accept an event every $6.25~{\rm ns}$.

\subsection{Performance}

\begin{figure}[htbp]
    \centering
    \includegraphics[width=0.90\columnwidth]{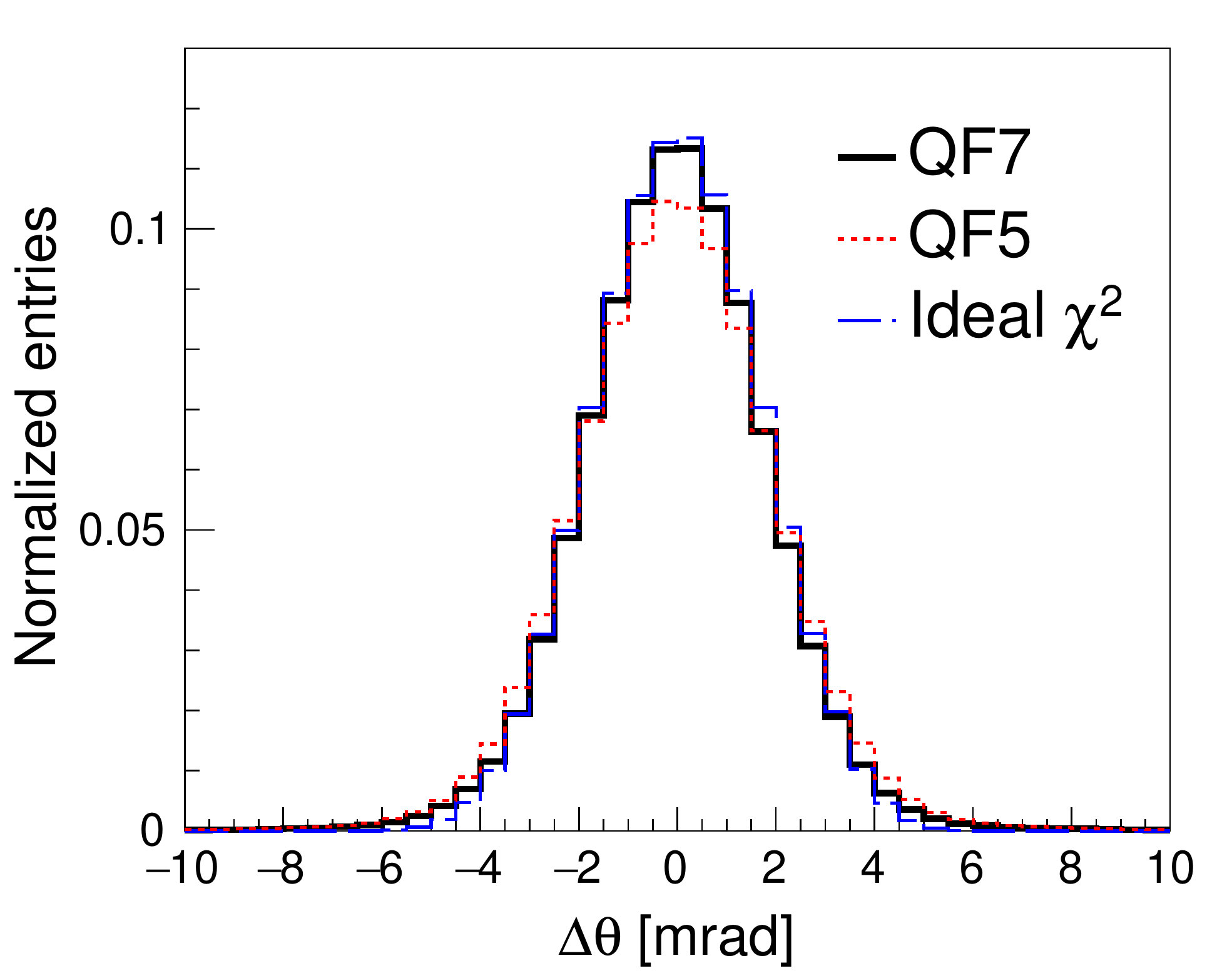}
    \caption{Distributions of the angle difference between the MC truth track segment
        and the reconstructed track segment.
        The solid (black) and dotted (red) histograms
        show the distributions for the track segments
        reconstructed by the \textbf{QF7} and \textbf{QF5} networks, respectively.
        The dashed (blue) histogram shows the distribution
        for track segments reconstructed by the ideal $\chi^2$ method.}
    \label{fig:rnn_f_comps}
\end{figure}

The angular resolution was evaluated from the widths of the distributions
of the difference between the network output and the MC truth angle.
Figure~\ref{fig:rnn_f_comps} shows the distributions for the \textbf{QF7} and \textbf{QF5} networks.
The distribution of the ideal $\chi^2$ method is also shown for reference.
Table~\ref{fig:perf_table} shows the values of the angular resolution
for the networks examined in this study.
The angular resolution of the \textbf{QF7},
\textbf{QF5}, and \textbf{QF3} networks ranges from 2.0 to 2.8~mrad.

\begin{table*}[htbp]
    \caption{The angular resolution, latency, and resource utilization
        for the neural networks optimized for the detector configuration of this study.
        The numbers in parentheses show
        the rates of resource utilization per super logic region.
        The angular resolution for \textbf{BL} is estimated using software,
        whereas those for \textbf{QF7}, \textbf{QF5}, and \textbf{QF3}
        are estimated using a Vivado post-implementation simulation.
        The \textbf{BL} network
        is provided without the intention to be run on the hardware;
        thus, the values of the latency and resource utilization are omitted.}
    \label{fig:perf_table}
    \centering
    \resizebox{\textwidth}{!}{
    \begin{tabular}{lcccccc}
        \hline
        Model        & Resolution~[mrad] & Latency~[ns] & DSP48        & LUT            & FF            & BRAM       \\
        \hline \hline
        \textbf{BL}  & $1.9$             & -            & -            & -              & -             & -          \\
        \textbf{QF7} & $2.0$             & 69           & 1389~(45\%) & 34,848~(8.1\%) & 5433~(0.6\%) & 75~(5.6\%) \\
        \textbf{QF5} & $2.2$             & 69           & 88~(2.9\%)   & 40,039~(9.3\%) & 3419~(0.4\%) & 75~(5.6\%) \\
        \textbf{QF3} & $2.8$           & 56 & 2~($<0.1$\%) & 21,682~(5.0\%) & 2242~(0.3\%) & 75~(5.6\%) \\
        \hline
    \end{tabular}
    }
\end{table*}

For the TGC detector at the ATLAS experiment at the HL-LHC,
an average angular resolution of 4 mrad is assumed~\cite{tdaq-tdr}.
The angular resolution depends on the detector region,
and the value obtained for $2.13 < |\eta| < 2.16$,
which covers the $\eta$ range of this study, is 2.4~mrad~\cite{public-plot}.
Although the detector configuration is not identical
and thus the numbers cannot be compared directly,
an angular resolution slightly greater than 2~mrad
achieved by the \textbf{QF7} and \textbf{QF5} networks
would be sufficient for the TGC detector.

The latency and the resource utilization were obtained for the quantized networks;
the results are summarized in Table~\ref{fig:perf_table}.
Each quantized network was implemented on Vivado
without timing violation
and achieved a latency less than 100~ns.
The resources were allocated by Vivado HLS
on the basis of the number of bits involved in the calculations;
some of multiplications performed by DSP48 in the \textbf{QF7} network
are performed by LUTs in the \textbf{QF5} and \textbf{QF3} networks.

The \textbf{QF7} network was trained
and evaluated on the datasets with artificial random noises.
The model was retrained on the noise levels of $10^{-4}$ and $10^{-3}$
and then implemented through Vivado.
The angular resolution of the model at each noise level
is shown in Table~\ref{perf_noise_table}.
With a reasonable noise level of $10^{-4}$,
the network could still outperform the iterative $\chi^2$ method;
it maintained comparable performance at a noise level of $10^{-3}$.

\begin{table}[htbp]
    \caption{Angular resolution in unit of milli-radians
    of the model trained on different datasets
    and implemented as \textbf{QF7}.
    The row headings are the noise level for the test sets,
    and the column headings are the noise level for the training sets.}
    \label{perf_noise_table}
    \centering
    \begin{tabular}{|c|ccc|}
        \hline
        \diagbox{Training}{Test} & 0   & $10^{-4}$ & $10^{-3}$ \\
        \hline
        0                        & 2.0       & 2.2       & 3.3       \\
        $10^{-4}$                & 2.0       & 2.1       & 2.6       \\
        $10^{-3}$                & 2.1       & 2.1       & 2.3       \\
        \hline
    \end{tabular}
\end{table}

\section{Conclusion}
\label{sec:conclusion}

Neural networks for fast tracking
with multiwire proportional chambers were proposed and studied.
A compact and quantized neural network
was developed with a conventional CNN structure,
where all the detector hits were inputted to a single convolution layer
followed by multilayer perceptrons.
A multistage neural network was then provided,
where the outputs of the convolution layers for gas-gap groups
were fed into multilayer perceptrons in sequence.
The multistage neural network demonstrated better
resource utilization and
shorter latency
for a given angular resolution of the tracking.
The firmware was implemented
in Xilinx UltraScale+ FPGA without timing violation
and provided an angular resolution comparable
with the best possible resolution of the detector
estimated by the $\chi^{2}$ method.
A trigger turn-on curve similar to or better than
that in Fig.~8.14 in Ref.~\cite{tdaq-tdr},
which was obtained using the $\chi^{2}$ method,
is expected with the neural networks developed in this study.
The performance degraded moderately and was still
acceptable with random noise hits.
The firmware was able to
perform one reconstruction in sub-$100$~ns with a fixed latency
and have a high throughput of $160$~MHz.

This work demonstrates a technique of forming a neural network
and reducing the required resources
for fast muon tracking using the hits of TGC-type detectors.
Although the present networks will need further modifications
and optimizations for a larger coverage of the detector
and for a full trigger chain, 
the techniques shown in this paper
can be exploited to maintain muon trigger thresholds
in the HL-LHC era despite the greater luminosity.

\section*{Acknowledgements}

This work was supported by JSPS KAKENHI Grant Numbers 16H06493, 18K03675, and 21H05085.

\bibliography{mybibfile}

\begin{thebibliography}{10}
\expandafter\ifx\csname url\endcsname\relax
  \def\url#1{\texttt{#1}}\fi
\expandafter\ifx\csname urlprefix\endcsname\relax\def\urlprefix{URL }\fi
\expandafter\ifx\csname href\endcsname\relax
  \def\href#1#2{#2} \def\path#1{#1}\fi

\bibitem{ATLAS_Higgs}
{ATLAS Collaboration}, {Observation of a new particle in the search for the
  Standard Model Higgs boson with the ATLAS detector at the LHC}, Physics
  Letters B 716~(1) (2012) 1--29.
\newblock \href
  {http://dx.doi.org/https://doi.org/10.1016/j.physletb.2012.08.020}
  {\path{doi:https://doi.org/10.1016/j.physletb.2012.08.020}}.

\bibitem{CMS_Higgs}
{CMS Collaboration}, {Observation of a new boson at a mass of 125 GeV with the
  CMS experiment at the LHC}, Physics Letters B 716~(1) (2012) 30--61.
\newblock \href
  {http://dx.doi.org/https://doi.org/10.1016/j.physletb.2012.08.021}
  {\path{doi:https://doi.org/10.1016/j.physletb.2012.08.021}}.

\bibitem{ATLAS_muon_trigger}
{ATLAS Collaboration}, {Performance of the ATLAS muon trigger in pp collisions
  at $\sqrt{s}=8$~TeV}, The European Physical Journal C 75 (2015) 120.
\newblock \href
  {http://dx.doi.org/https://doi.org/10.1140/epjc/s10052-015-3325-9}
  {\path{doi:https://doi.org/10.1140/epjc/s10052-015-3325-9}}.

\bibitem{CMS_l1_trigger}
{CMS Collaboration}, {The CMS trigger system}, Journal of Instrumentation
  12~(01) (2017) P01020.
\newblock \href
  {http://dx.doi.org/https://doi.org/10.1088/1748-0221/12/01/p01020}
  {\path{doi:https://doi.org/10.1088/1748-0221/12/01/p01020}}.

\bibitem{ATLAS_Higgs10}
{ATLAS Collaboration}, {A detailed map of Higgs boson interactions by the ATLAS
  experiment ten years after the discovery}, Nature 607 (2022) 52--59.
\newblock \href {http://dx.doi.org/https://doi.org/10.1038/s41586-022-04893-w}
  {\path{doi:https://doi.org/10.1038/s41586-022-04893-w}}.

\bibitem{CMS_Higgs10}
{CMS Collaboration}, {A portrait of the Higgs boson by the CMS experiment ten
  years after the discovery}, Nature 607 (2022) 60--68.
\newblock \href {http://dx.doi.org/https://doi.org/10.1038/s41586-022-04892-x}
  {\path{doi:https://doi.org/10.1038/s41586-022-04892-x}}.

\bibitem{hl-lhc}
I.~B. Alonso, O.~Br\"uning, P.~Fessia, M.~Lamont, L.~Rossi, L.~Tavian,
  M.~Zerlauth, {High-Luminosity Large Hadron Collider (HL-LHC): Technical
  design report} (2020).
\newblock \href {http://dx.doi.org/https://doi.org/10.23731/CYRM-2020-0010}
  {\path{doi:https://doi.org/10.23731/CYRM-2020-0010}}.

\bibitem{tdaq-tdr}
{ATLAS Collaboration}, \href{https://cds.cern.ch/record/2285584}{{Technical
  Design Report for the Phase-II Upgrade of the ATLAS TDAQ System}} (2018).
\newline\urlprefix\url{https://cds.cern.ch/record/2285584}

\bibitem{ML_summary}
\href{https://iml-wg.github.io/HEPML-LivingReview/}{{A Living Review of Machine
  Learning for Particle Physics}}.
\newline\urlprefix\url{https://iml-wg.github.io/HEPML-LivingReview/}

\bibitem{Duarte_2018}
J.~Duarte, S.~Han, P.~Harris, S.~Jindariani, E.~Kreinar, B.~Kreis, J.~Ngadiuba,
  M.~Pierini, R.~Rivera, N.~Tran, Z.~Wu, {Fast inference of deep neural
  networks in FPGAs for particle physics}, Journal of Instrumentation 13~(07)
  (2018) P07027.
\newblock \href
  {http://dx.doi.org/https://doi.org/10.1088/1748-0221/13/07/p07027}
  {\path{doi:https://doi.org/10.1088/1748-0221/13/07/p07027}}.

\bibitem{Summers_2020}
S.~Summers, G.~D. Guglielmo, J.~Duarte, P.~Harris, D.~Hoang, S.~Jindariani,
  E.~Kreinar, V.~Loncar, J.~Ngadiuba, M.~Pierini, D.~Rankin, N.~Tran, Z.~Wu,
  {Fast inference of Boosted Decision Trees in FPGAs for particle physics},
  Journal of Instrumentation 15~(05) (2020) P05026.
\newblock \href
  {http://dx.doi.org/https://doi.org/10.1088/1748-0221/15/05/p05026}
  {\path{doi:https://doi.org/10.1088/1748-0221/15/05/p05026}}.

\bibitem{Ngadiuba_2020}
J.~Ngadiuba, V.~Loncar, M.~Pierini, S.~Summers, G.~D. Guglielmo, J.~Duarte,
  P.~Harris, D.~Rankin, S.~Jindariani, M.~Liu, K.~Pedro, N.~Tran, E.~Kreinar,
  S.~Sagear, Z.~Wu, D.~Hoang, {Compressing deep neural networks on FPGAs to
  binary and ternary precision with hls4ml}, Machine Learning: Science and
  Technology 2~(1) (2020) 015001.
\newblock \href {http://dx.doi.org/https://doi.org/10.1088/2632-2153/aba042}
  {\path{doi:https://doi.org/10.1088/2632-2153/aba042}}.

\bibitem{qkeras}
C.~N. {Coelho Jr}, A.~Kuusela, S.~Li, H.~Zhuang, J.~Ngadiuba, T.~K. Aarrestad,
  V.~Loncar, M.~Pierini, A.~A. Pol, S.~Summers, {Automatic heterogeneous
  quantization of deep neural networks for low-latency inference on the edge
  for particle detectors}, Nature Machine Intelligence 3 (2021) 675--686.
\newblock \href {http://dx.doi.org/https://doi.org/10.1038/s42256-021-00356-5}
  {\path{doi:https://doi.org/10.1038/s42256-021-00356-5}}.

\bibitem{CMS-PhaseII-L1-TDR}
{CMS Collaboration}, \href{https://cds.cern.ch/record/2714892}{{The Phase-2
  Upgrade of the CMS Level-1 Trigger}} (2020).
\newline\urlprefix\url{https://cds.cern.ch/record/2714892}

\bibitem{FTK}
{ATLAS Collaboration}, {The ATLAS Fast TracKer system}, Journal of
  Instrumentation 16~(07) (2021) P07006.
\newblock \href
  {http://dx.doi.org/https://doi.org/10.1088/1748-0221/16/07/P07006}
  {\path{doi:https://doi.org/10.1088/1748-0221/16/07/P07006}}.

\bibitem{CDF_SVT}
W.~Ashmanskas, A.~Bardi, M.~Bari, S.~Belforte, J.~Berryhill, M.~Bogdan,
  A.~Cerri, A.~Clark, G.~Chlanchidze, R.~Condorelli, R.~Culbertson,
  M.~Dell'Orso, S.~Donati, H.~Frisch, S.~Galeotti, P.~Giannetti, V.~Glagolev,
  A.~Leger, E.~Meschi, F.~Morsani, T.~Nakaya, G.~Punzi, L.~Ristori, H.~Sanders,
  A.~Semenov, G.~Signorelli, M.~Shochet, T.~Speer, F.~Spinella, P.~Wilson,
  X.~Wu, A.~Zanetti,
  \href{https://www.sciencedirect.com/science/article/pii/S016890020000190X}{{Silicon
  vertex tracker: a fast precise tracking trigger for CDF}}, Nuclear
  Instruments and Methods in Physics Research Section A: Accelerators,
  Spectrometers, Detectors and Associated Equipment 447~(1) (2000) 218--222.
\newblock \href
  {http://dx.doi.org/https://doi.org/10.1016/S0168-9002(00)00190-X}
  {\path{doi:https://doi.org/10.1016/S0168-9002(00)00190-X}}.
\newline\urlprefix\url{https://www.sciencedirect.com/science/article/pii/S016890020000190X}

\bibitem{ML_muon_CMS}
M.~Migliorini, J.~Pazzini, A.~Triossi, M.~Zanetti, A.~Zucchetta,
  \href{https://arxiv.org/abs/2105.04428}{{Muon trigger with fast Neural
  Networks on FPGA, a demonstrator}} (2021).
\newblock \href {http://arxiv.org/abs/2105.04428 [hep-ex]}
  {\path{arXiv:2105.04428 [hep-ex]}}.
\newline\urlprefix\url{https://arxiv.org/abs/2105.04428}

\bibitem{vivado}
Xilinx, \href{https://www.xilinx.com}{{Vivado Design Suite}}.
\newline\urlprefix\url{https://www.xilinx.com}

\bibitem{G4-short}
S.~Agostinelli, et~al., {Geant4--a simulation toolkit}, Nuclear Instruments and
  Methods in Physics Research Section A: Accelerators, Spectrometers, Detectors
  and Associated Equipment 506~(3) (2003) 250--303.
\newblock \href
  {http://dx.doi.org/https://doi.org/10.1016/S0168-9002(03)01368-8}
  {\path{doi:https://doi.org/10.1016/S0168-9002(03)01368-8}}.

\bibitem{tgc_structure}
{ATLAS Collaboration}, {The ATLAS Experiment at the CERN Large Hadron
  Collider}, Journal of Instrumentation 3~(08) (2008) S08003.
\newblock \href
  {http://dx.doi.org/https://doi.org/10.1088/1748-0221/3/08/s08003}
  {\path{doi:https://doi.org/10.1088/1748-0221/3/08/s08003}}.

\bibitem{keras}
F.~Chollet, {\it et al.}, \href{https://github.com/fchollet/keras}{{Keras}}
  (2015).
\newline\urlprefix\url{https://github.com/fchollet/keras}

\bibitem{tensorflow2015-whitepaper}
M.~Abadi, A.~Agarwal, P.~Barham, E.~Brevdo, Z.~Chen, C.~Citro, G.~S. Corrado,
  A.~Davis, J.~Dean, M.~Devin, S.~Ghemawat, I.~Goodfellow, A.~Harp, G.~Irving,
  M.~Isard, Y.~Jia, R.~Jozefowicz, L.~Kaiser, M.~Kudlur, J.~Levenberg,
  D.~Man\'{e}, R.~Monga, S.~Moore, D.~Murray, C.~Olah, M.~Schuster, J.~Shlens,
  B.~Steiner, I.~Sutskever, K.~Talwar, P.~Tucker, V.~Vanhoucke, V.~Vasudevan,
  F.~Vi\'{e}gas, O.~Vinyals, P.~Warden, M.~Wattenberg, M.~Wicke, Y.~Yu,
  X.~Zheng, \href{https://www.tensorflow.org/}{{TensorFlow: Large-Scale Machine
  Learning on Heterogeneous Systems}}, software available from tensorflow.org
  (2015).
\newline\urlprefix\url{https://www.tensorflow.org/}

\bibitem{vivado_hls}
Xilinx,
  \href{https://www.xilinx.com/support/documentation/sw_manuals/xilinx2020_1/ug902-vivado-high-level-synthesis.pdf}{{Vivado
  Design Suite User Guide: High-Level Synthesis}} (2021).
\newline\urlprefix\url{https://www.xilinx.com/support/documentation/sw_manuals/xilinx2020_1/ug902-vivado-high-level-synthesis.pdf}

\bibitem{VGG}
K.~Simonyan, A.~Zisserman, \href{https://arxiv.org/abs/1409.1556}{{Very deep
  convolutional networks for large-scale image recognition}} (2014).
\newblock \href {http://arxiv.org/abs/1409.1556 [cs.CV]} {\path{arXiv:1409.1556
  [cs.CV]}}.
\newline\urlprefix\url{https://arxiv.org/abs/1409.1556}

\bibitem{Hennequin_2020}
A.~Hennequin, B.~Couturier, V.~Gligorov, S.~Ponce, R.~Quagliani, L.~Lacassagne,
  A fast and efficient {SIMD} track reconstruction algorithm for the {LHCb}
  upgrade 1 {VELO}-{PIX} detector, Journal of Instrumentation 15~(06) (2020)
  P06018.
\newblock \href
  {http://dx.doi.org/https://doi.org/10.1088/1748-0221/15/06/p06018}
  {\path{doi:https://doi.org/10.1088/1748-0221/15/06/p06018}}.

\bibitem{public-plot}
{ATLAS Collaboration},
  \href{https://twiki.cern.ch/twiki/bin/view/AtlasPublic/L0MuonTriggerPublicResults}{{L0MuonTriggerPublicResults}}.
\newline\urlprefix\url{https://twiki.cern.ch/twiki/bin/view/AtlasPublic/L0MuonTriggerPublicResults}

\end{thebibliography}

\end{document}